\documentclass[conference]{IEEEtran}

\usepackage{cite}
\usepackage[cmex10]{amsmath}
\usepackage{comment}
\usepackage{xcolor}
\usepackage{algorithmic}
\usepackage{array}  
\usepackage{mdwmath}
\usepackage{mdwtab}
\usepackage{eqparbox}
\usepackage[tight,footnotesize]{subfigure}
\usepackage{booktabs}
\usepackage{url}
\usepackage[colorlinks]{hyperref}
\usepackage[obeyspaces]{xurl}
\newcommand{\cut}[1]{}

\ifCLASSINFOpdf
  \usepackage[pdftex]{graphicx}
  \graphicspath{{./imgs/}}
\else
  \usepackage[dvips]{graphicx}
\fi

\begin{document}
%
\title{On the Cyber-Physical Security of Commercial Indoor Delivery Robot Systems
\vspace{-0.2in}}


\author{\IEEEauthorblockN{
Fayzah Alshammari, 
Yunpeng Luo and Qi Alfred Chen}
\IEEEauthorblockA{Univesity of California, Irvine
}
}

\IEEEoverridecommandlockouts

\maketitle

\begin{abstract}
Indoor Delivery Robots (IDRs) play a vital role in the upcoming fourth industrial revolution, autonomously navigating and transporting items within indoor environments. In this work, we thus aim to conduct the first security analysis of the IDR systems considering both cyber- and physical-layer attack surface and domain-specific attack goals across security, safety, and privacy. As initial results, we formulated a general IDR system architecture from 40 commercial IDR models and then performed an initial cyber-physical attack entry point identification. We also performed an experimental analysis of a real commercial IDR robot-side software and identified several vulnerabilities. We then discuss future steps.


\end{abstract}

\section{Introduction}
\label{sec: intro}
Indoor Delivery Robots (IDRs) are crucial participants in the upcoming fourth industrial revolution, autonomously navigating and transporting items within indoor environments. The global market for IDRs was valued at USD 6.106 Million in 2020 and is projected to reach USD 157.618 Million by 2027. Furthermore, IDRs are complex automation devices that can interact with people through multiple control systems in public places such as hospitals, hotels, restaurants, and airports. They can also interact with other indoor existing systems, such as facility management systems, and remotely with vendors for support and maintenance. This brings critical cyber security, physical security, and safety concerns. To address these concerns and to ensure the security of IDRs and, more importantly, the safety of people around them, it is vital to understand the potential security challenges that assist in developing robust security strategies for IDR systems. Various prior works are studying the security of robotic systems generally such as as~\cite{salt_to_pepper}, where they mainly report vulnerabilities in network surface attacks, but no comprehensive security analysis of the indoor delivery robot has been done.

In this work, we thus aim to conduct the first comprehensive security analysis of commercial IDRs, addressing a gap in systematic evaluations of their attack surface and cyber-physical impacts. 
We aim to consider cyber- and physical-layer threats in our threat model and identify 
domain-specific attack goals across security, safety, and
privacy. As an initial step, we survey the market of commercial IDRs, coming up with a list of 40 commercial models. We then 
utilize this list to derive the system architecture for commercial IDRs to identify attack entry points, and also perform an experimental analysis of a commercial IDR robot-side software. In the future, we will perform further domain-specific attack class discovery, experimental analysis, and defense discussions. 


\section{Threat Model and Problem Formulation}
\textbf{Threat model.} This work assumes local-environment attackers when carrying out physical-layer attacks or cybercriminals when conducting cyber-layer attacks. The attacker is assumed to possess the required skills to reverse engineer the IDR's software and know the target delivery robot's structure. This can be realistically achieved by purchasing the same IDR model herself or renting one.

\textbf{Attack goals.} By targeting an IDR, an attacker can achieve various domain-specific security, safety, and privacy breaches, for example: (1) \textit{delivery service manipulation} which includes altering the destination, replacing the delivered item with harmful or unexpected substances, or simply preventing the delivery from happening at all, (2) \textit{safety damages}, including crashing into people, and other robots, and (3) \textit{sensitive information stealing} such as information about the indoor environment’s maps, surroundings and people. 

\section{Analysis Methodology}
\label{sec:attack_method}
With the threat model and attackers' goals described above, we plan to systematically analyze the security of the state-of-the-art commercial IDRs by the following steps:

\begin{itemize}
    \item \textbf{Comprehensive attack surface analysis}. To perform a comprehensive attack surface analysis of IDR systems, we must first obtain a general IDR system architecture that can comprehensively describe their state-of-the-art system components and interaction designs in indoor environments. To achieve this, we exhaustively search for all IDR models from the companies dominating the IDR market today and then use their publicly available information (e.g., user manuals) to derive such a general IDR system architecture. From this derived system architecture, we can then comprehensively identify potential attack entry points at both the cyber- and physical-layers;
    
    \item \textbf{Domain-specific attack class discovery}. Based on the system architecture, system requirements, and attack entry points, we plan then to systematically identify concrete IDR-specific attack classes driven by the domain-specific attack goals stated earlier;
    
    \item \textbf{Experimental security analysis of a real IDR system}. To understand the feasibility of these identified domain-specific attack classes, we plan to perform an experimental security analysis of real IDR systems and components guided by the discovered attack classes above. We will finally discuss potential defence strategies based on the insights from these analyses.
\end{itemize}

\section{Evaluation}
\textbf{Testing Environment} 
We obtained the robot-side software of one of the commercial IDRs, which we examined and tested in an environment that encompassed both Ubuntu 22.04 and Windows 10 operating systems. To facilitate our evaluation process, we utilized Genymotion as an emulator, enabling the creation of a virtual Android environment wherein the robot's application could be executed. Furthermore, we leveraged MOBSF, an open-source framework designed for mobile application security analysis, to systematically identify vulnerabilities within the application being tested. We then use Jadx to reverse-engineer the application APK files. Finally, we utilized APKtool to modify and rebuild the APK as needed.

\section{Early Results}
\subsection{IDRs List and Information Gathering}
\begin{figure} [t]
    \centering
    \includegraphics[width=0.8\columnwidth]{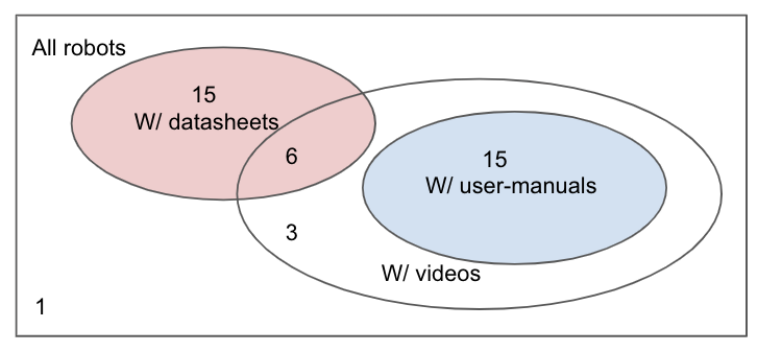}
    \vspace*{-3mm}
    \caption{Venn diagram for the used information sources for deriving the general IDR system architecture. As shown, 39 of 40 IDRs are covered by at least 1 type of information source (i.e., dataset, user manual, or video).}
    \label{fig:venn}
    \vspace*{-6mm}
\end{figure}

To establish the IDR system architecture, we identified 14 leading companies from recent market
reports and manually reviewed their websites for IDR products available between April 3 and June 7, 2023. This included all robots designed for indoor delivery in environments like hospitals and restaurants, resulting in a list of 40 IDR models. We collected information on those models using publicly-available datasheets, user manuals, and videos as visualized in Fig.~\ref{fig:venn}. Specifically, datasheets were found for 21 models, which provided technical specifications; user manuals for 15 models, which offered operational details; and videos for 24 models, which further helped to understand robot functionality and interactions. In the following, we provide additional information about the companies and the indoor delivery robots (IDRs) we considered, along with the process of gathering and processing data about them.

\textbf{Companies.} We utilize commercial marketing reports such as Research and Market~\cite{global_report}, Market and Markets~\cite{market} and the UCI marketing database~\cite{uci_market} and the online market for robots such as The Robot ~\cite{robot_marketplace} and Alibaba~\cite{ali}. We particularly used the prominent players in the market of indoor delivery robots from these reports: Global Indoor Delivery Robot Report (Aug 2021) and Forecast 2027 and Opportunities of Robotics in Industry~\cite{opport_report}. Upon completing our market survey, we have a list comprising 14 robot companies.

\textbf{IDRs models.} We proceeded to examine each company's website manually to identify the range of available products, specifically focusing on indoor delivery robots designed for environments such as hospitals, hotels, and restaurants. This exploration led us to compile a list of 40 delivery robots, which we subsequently categorized into two distinct types: in-floor robots and in-building robots. In-floor robots can navigate within a single floor while in-building robots demonstrate the ability to traverse multiple floors within a building, facilitated by their interaction with elevators.

\textbf{Robots Information Gathering and Processing}
We then used three main data sources to collect information about these products: datasheets, user manuals and videos. Datasheets refer to the marketing and high-level technical specifications, characteristics and functionality of specific robots, such as the robot type, speed, weight, shape, connectivity, localisation techniques and in some cases the robot’s OS. These sheets usually are publicly available on the vendor's website and we found 21 sheets. User manuals, on the other hand, are the operational and user guides that usually come with the robot to help the end user to unpackage, set up, and operate the robot. It also has some debugging and safety-relevant information.

We used two publicly available websites that host several product manuals: device.report~\cite{device_report} and manualslib.com~\cite{manuals_lib}. We were able to find 15 user manuals, two of them are directly requested from the vendor. Finally, we also used videos when available, some of them hosted on the vendor’s website and in most cases available on YouTube. The videos help with learning more about the functionality and the general workflow of the robot while in action. It also helps us learn more about the interaction between robots, humans and the indoor environment as well.

\subsection{General IDR System Architecture} 
\begin{figure*}[ht]
\centering
\includegraphics[width=0.95\textwidth]{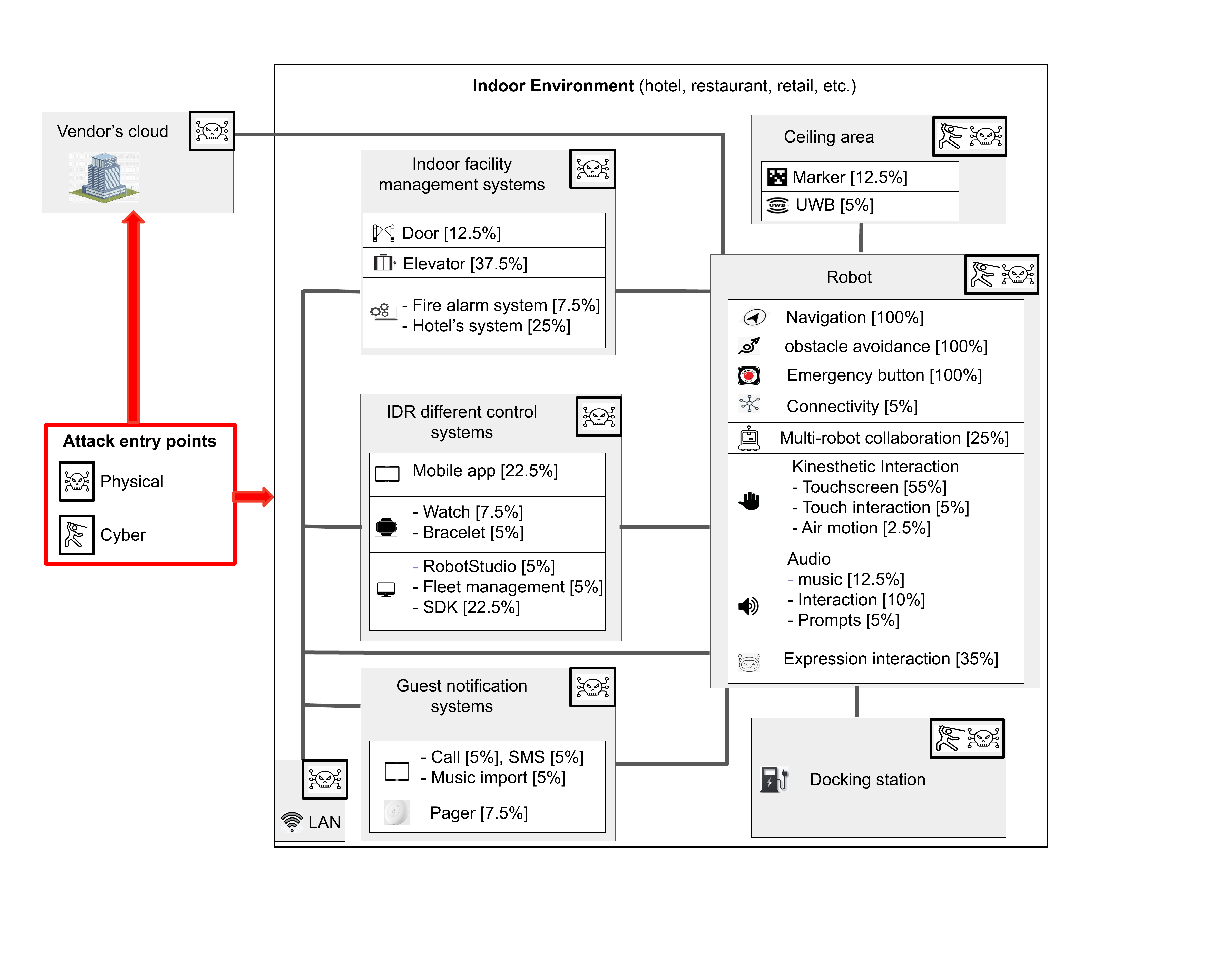} 
\vspace{-19mm} 
\caption{General Indoor Delivery Robot (IDR) system architecture and an initial cyber-physical attack entry point analysis, where [\%] denotes the percentage of robots with a specific feature/component. 
}
\label{fig:system-structure}
\vspace{-6mm} 
\end{figure*}

 We then use the gathered information above to derive a general system architecture of today's commercial IDR systems, which involves various types of domain-specific system components such as door/evaluator control, multi-robot collaboration, phone/watch robot calls, pager, markers on the ceiling, etc. as shown in Fig.~\ref{fig:system-structure}. The diagram presents a detailed overview of the IDR system's architecture, highlighting its integration with facility management systems like elevators, a heavy reliance on the vendor's cloud (30\%), and diverse control systems like mobile apps and wearables, reflecting operational flexibility and user preference accommodation. Robot features such as navigation, obstacle avoidance, and emergency stop buttons are standard in all robots despite some limitations in robot collaboration capabilities. Touchscreens and expression interaction emphasize user-friendliness, and guest notification systems focus on enhancing service efficiency and guest interaction.

\begin{itemize}
    \item \textbf{Ceiling area.} This is crucial in supporting indoor delivery robots by hosting navigational aids like markers and Ultra-Wideband (UWB) technology. These tools enable robots to orient themselves, navigate precisely within the space, and execute tasks efficiently. Markers provide fixed reference points for the robots, while UWB offers real-time, accurate location data, essential for navigating through complex indoor environments and ensuring accurate deliveries.

    \item \textbf{IDR.} An indoor delivery robot is a specialized robot designed to autonomously navigate indoor environments and deliver items from one location to another.  Based on ISO 8373:2021: this type of robot is a mobile service robot (not an industrial robot) that travels autonomously and is used professionally to perform delivery tasks~\cite{iso_def}. Indoor delivery robots use a combination of sensors and mapping techniques to navigate their surroundings. They may employ sensors like cameras, laser scanners, ultrasonic sensors, or depth sensors to detect and avoid obstacles such as walls, furniture, and/or people.
    
    The robot may also create and update a map of the environment to plan its path effectively. The robot may use techniques such as simultaneous localization and mapping (SLAM) or utilize markers or beacons placed in the environment for reference to determine its position within the environment.

    \item \textbf{Docking station.}The primary function of a docking station is to recharge the robot's batteries. It serves as a power source where robots autonomously return when their energy levels are low, ensuring they have minimal downtime and are ready for continuous operation.

    \item \textbf{Indoor facility management systems.}This includes robot interaction with markers, e-doors, elevators, and other exciting systems in the indoor environment, such as the Fire Alarm System and the hotel’s system.
    
    \item \textbf{IDR control systems.} This component manages and coordinates indoor delivery robots, including mobile apps for setting delivery destinations, tracking robots, and receiving updates, enhancing user interaction with the system. It also includes wearable device integration, such as watches and bracelets, enabling on-the-go notifications and control. With software integrations like RobotStudio for robot programming and fleet management, the control system can be customized.  
    
    \item \textbf{Guest notification systems.} This includes multiple user interfaces, such as touch screens or voice command systems, to interact with end users. They may also use audio or visual signals to indicate their status or intentions to nearby people in the environment. 
    
    \item \textbf{Vendor's cloud.} This includes the interaction between the robot and the vendor of the robot for support, tracking, and remote controlling purposes.   
\end{itemize} 

\subsection{Initial Security Analysis}

Based on the derived general IDR system architecture, we perform an initial attack surface analysis for each IDR system component by analyzing their potential cyber- and physical-attack entry points. Specifically, for \textit{cyber-attack entry points}, components are categorized based on their reliance on software or firmware, like control systems and communication protocols; their need for network connectivity, which includes elements like LAN and cloud services; their involvement in data storage, retrieval, or transmission, which can be vulnerable to breaches and unauthorized access; and the ability for remote control or access, such as mobile apps or web interfaces, due to hacking risks. On the other hand, \textit{physical-attack entry points} are defined by hardware-based components, including sensors, cameras, and ports; components requiring physical interaction or vulnerable to tampering, like touchscreens and docking stations; elements dependent on the physical environment, such as navigation systems; and manual safety controls like emergency stop buttons, which can be directly and manually interfered with. Such initial analysis results are shown in Fig.~\ref{fig:system-structure}.



Meanwhile, we obtained the robot-side software of one of the commercial IDRs we examined.
Since we are in the process of planning for a responsible vulnerability disclosure to the company for this IDR, we anonymize the application's name in this version. This software is an Android app, and through static penetration testing, we uncovered two types of software vulnerabilities based on the OWASP Mobile Top 10 vulnerability categorizations~\cite{owaspm}. The first, \textit{Insufficient Binary Protection}, enabled us to reverse-engineer the app, exposing crucial details about its functionality. This vulnerability also allowed us to modify the app's code due to the absence of code tampering detection or protection mechanisms. An example is our successful manipulation of the app's user interface, demonstrating the potential for altering IDR functionalities, such as delivery details. The second vulnerability pertains to \textit{Insecure Data Storage}, as we discovered that the app's \texttt{android:allowBackup} attribute is set to \texttt{True}. This setting permits the creation of backups that include application data, which could lead to the theft of sensitive information such as indoor maps, delivery schedules, and customer details. In the following section, more details about the vulnerabilities discovered will be provided. 


\textbf{Lack of Code Obfuscation}
Our analysis revealed that the application's binary is not obfuscated, enabling clear visibility and analysis of the Java code within the Android application package (APK). This lack of obfuscation allows an attacker to extract the APK from a local Android device and decompile the binary, unveiling the underlying source code logic. Moreover, an attacker could alter the binary to circumvent business logic or modify configuration settings, then re-sign and reinstall the app. For instance, the login functionality code was found to be directly accessible and readable in its decompiled form. A CVE submission is planned for this vulnerability to ensure it is properly documented and addressed within the security community.


\textbf{Code Tampering}
The application is missing crucial mechanisms such as checksum validation, code signing verification, and runtime integrity checks that detect or prevent code tampering. Without these safeguards, attackers can modify the app's code and behavior, compromising the principles of secure design. This vulnerability exposes the application to unauthorized alterations, such as changing its functionality or user interface. For example, during our analysis, we successfully modified the application’s user interface, demonstrating how the absence of integrity checks allows for unauthorized changes. A CVE submission is planned to document this vulnerability as well.


\textbf{Insecure Data Backup Configuration}
The application's setting of the android:allowBackup attribute to True presents a critical security flaw. This configuration permits attackers to access the backup data without authorization, posing a significant risk of compromising the application, its users, and any related services. The vulnerability could facilitate various malicious activities, such as data theft, user impersonation, or exploitation of service weaknesses. During our analysis, we identified this misconfiguration in the application, where the android:allowBackup attribute was set to True. This allows attackers to extract sensitive backup data and potentially use it to compromise the application, users, or associated services. A CVE submission is planned to formally document this vulnerability in adherence to the responsible vulnerability disclosure practices. 


This section delves into the existing literature related to the security experimental analysis of commercial robots. While several studies, such as ~\cite{salt_to_pepper}~\cite{to_make_robot_secure}, have investigated vulnerabilities in various research robots, focusing primarily on network surface attacks, a comprehensive security analysis specifically targeting indoor delivery robots remains absent. Notably, one study~\cite{industrail}, closely aligns with our research, conducting an experimental security analysis on industrial robots. However, this study falls short of addressing non-industrial robots. Our research aims to fill this gap by providing a thorough security analysis of indoor delivery robots, expanding the understanding of their security landscape.

\section{Conclusion and Future Plans}
In this work, we aim to perform the first security analysis of IDR systems that encompass both cyber and physical layers and target specific attack goals related to security, safety, and privacy. As an initial result, we formulated a general IDR system architecture based on 40 commercial IDR models today and then performed an initial cyber-physical attack surface analysis. We also performed an initial experimental security evaluation of an actual commercial IDR software, yielding two critical vulnerabilities. In the future, we plan to (1) build upon the derived general IDR system architecture and the concrete cyber-physical attack surface analysis to start the IDR-specific attack class discovery and (2) extend the security analysis to a complete IDR system (not just the robot-side software).

\section*{Acknowledgements}
This research was supported in part by NSF under grants CNS2145493 and USDOT under Grant 69A3552348327 for the CARMEN+ UTC.

\bibliographystyle{IEEEtran}
\bibliography{main.bib}
\end{document}